\documentclass[preprint,prd,aps,showpacs]{revtex4}
\usepackage{latexsym}
\usepackage{slashed}
\usepackage{graphicx}
\usepackage{epsfig}
\usepackage{epstopdf}
\usepackage{color}
\usepackage{amsmath}
\begin{document}
\title{Cancellation of Infrared divergences to all orders in LFQED}
\author{Jai D. More\footnote{\tt more.physics@gmail.com}}
\author{Anuradha Misra\footnote{\tt misra@physics.mu.ac.in}}
\affiliation{Department of Physics,University of Mumbai,
\\Santa Cruz(E), Mumbai, India-400098}%
\begin{abstract}
We present an all order proof  of cancellation of infrared (IR) divergences in Light Front Quantum Electrodynamics (LFQED) using a coherent state approach. It is shown,  using fermion mass renormalization as an example, that the true IR divergences  are eliminated to all orders in a light front time ordered perturbative calculation  if one uses CSB instead of the usual Fock basis to calculate the  Hamiltonian matrix elements.
\end{abstract}
\date{\today}
\maketitle 
\section{Introduction}

Light Front Field Theories (LFFT's) have emerged as a strong candidate amongst theoretical formalisms that can 
provide an answer to  the yet unresolved problem of a relativistic bound state system. If the present attempts  to apply the methods developed in this area of research   
are successful, one will be able to obtain,  starting form basic principles, the wave functions of hadrons  which are important input 
in the calculation of hadronic cross sections \cite{WILSON94, BRODSKY97}.  An important issue that needs to be understood 
for the success of these efforts is the problem of infrared (IR) divergences which have different structure from those in equal time theory.  In this work, we address the 
important issue of IR cancellation in light front quantum electrodynamics (LFQED).

It is well known that the infrared divergences in equal--time formulation of quantum electrodynamics (QED) are eliminated to all orders due to a cancellation between the real and virtual 
contributions by virtue of the famous  
Bloch and Nordseick Theorem\cite{BLOCH37}. The theorem is based on the idea that in an actual experiment involving charged particles, one cannot  specify 
the final state completely as due to the finite size of the detector, the charged particle can be accompanied by any number of soft photons. Therefore, in cross section calculations, one must sum over all possible final states 
taking into account emission of soft real photons which might have escaped detection. The Bloch-Nordseick mechanism takes into account all states with any number 
of soft photons below experimental resolution thus leading to cancellation of IR divergences. 

In QED, a general approach to treat IR divergences was given by Yennie {\it etal}~\cite{YENNIE61}. In this approach, IR divergences
are factored out and then treated to all orders of covariant perturbation theory to give a residual perturbative expansion which is 
IR finite. These IR factors are then expressed in exponential form leading to cancellation of IR divergences between the  real and virtual 
photon contributions. These factors depend only on the external momenta of charged particles and are independent of
the momentum of the intermediate interaction terms.

Following the work of Yennie {\it etal}, Chung \cite{CHU65} showed that IR divergences indeed cancel to all orders in perturbation theory 
at the level of amplitude itself provided the initial and final states are chosen properly. The condition for this cancellation constrains the initial and 
final states, which are actually  charged particles with a superposition of an infinitely large number of soft photons, to belong to a new space instead of the usual Fock space. 

Kulish and Faddeev (KF)\cite{KUL70} developed the method of asymptotic dynamics and obtained a set of 
asymptotic states starting with a modified, relativistically and gauge invariant definition of S--matrix and showed the cancellation of IR divergences at amplitude level using this basis.  

 KF were the first to show that in QED, the asymptotic Hamiltonian does not coincide with the free Hamiltonian. They  constructed the asymptotic Hamiltonian $V_{as}$ for QED thus modifying the asymptotic condition to introduce a new space of asymptotic states given by
\begin{displaymath}
\vert n;\pm \rangle = \Omega^A_{\pm}\vert n\rangle 
\end{displaymath}
where $\Omega^A_{\pm} $  defined by
\begin{displaymath}
\Omega^A_{\pm} = T~exp\bigg[-i\int_{\mp\infty}^0 V_{as}(t) dt \bigg]
\end{displaymath}
is the asymptotic evolution operator and $\vert n\rangle$ is the Fock state.
The transition matrix elements formed by using these states are IR finite.

KF approach was applied by Greco etal\cite{GRECO78} to study the IR behavior  of  non abelian gauge theories using coherent states of definite color and factorized in  fixed angle regime. The matrix elements using these coherent states were shown to be IR finite, first to the lowest order and then to all orders  under the condition that  the soft meson formula for real gluons holds to all orders.

 Coherent states in the context of  Light Front Field Theory (LFFT) have  been discussed by various authors\cite{Vary1988, Vary1999}. A coherent state approach has been developed by one of us and applied to show the cancellation 
of true IR divergences in one loop vertex correction in LFQED in Ref.~\cite{ANU94}. Subsequently, the formalism was applied to show the same in LFQCD \cite{ANU00}. 
Possibility of practical application of the method was discussed in Ref.~\cite{ANU96} where the coherent state method was applied to obtain an IR divergence free light-cone Schr\"odinger equation for positronium. 

It has also been shown that the IR divergences in fermion self energy correction at two loop order cancel in LFQED \cite{JAI12, JAI13} if one uses  coherent state approach. 
%In this approach cancellation of IR divergences has been found to occur separately in each set of diagrams \cite{ANU94, ANU00, JAI12, JAI13}. 
% Thus IR divergences cancels to the corresponding coherent state diagrams. 

In this work, we  demonstrate the cancellation of IR divergences in 
fermion mass renormalization in Light Front Quantum Electrodynamics (LFQED) to all orders by using a 
coherent state basis (CSB) for  calculating  the Hamiltonian matrix elements. But till date no such proof exists in LFQCD due to the confinement property of QCD. 

 In the following sections, we shall present a most general and a rather formal all order proof of cancellation of IR divergences in fermion mass renormalization in LFQED  by using a CSB for calculating the Hamiltonian matrix elements.
\section{IR Divergences and Coherent State Formalism} 
The coherent state method is based on the observation that for theories with long range interactions or theories  having bound states as asymptotic states, the total Hamiltonian does not reduce to the free field Hamiltonian in the limit $\vert t \vert \rightarrow \infty$. In LFFT, the asymptotic Hamiltonian $H_{as}$,  is evaluated by taking the limit $\vert x^+\vert\rightarrow \infty$ of the interaction Hamiltonian. Each term in the interaction Hamiltonian 
$H_{int}$ has a light-cone time dependence of the form  $exp[-i(p_1^-+p_2^-+\cdots+p_n^-)x^+]$ and therefore, if  $(p_1^-+p_2^-+\cdots+p_n^-)$ vanishes at  some vertex, then the corresponding term in $H_{int}$ does not vanish in large $x^+$ limit. Thus, the total Hamiltonian can be written as 
\begin{equation}
H = H_{as} + H_I^\prime
\end{equation}
where 
\begin{equation}
H_{as}(x^+) = H_0 +V_{as}(x^+)
\end{equation}
The associated $x^+$ evolution operator $U_{as}(x^+) $ in the Schr\"odinger representation, which satisfies the equation 
\begin{equation} 
i\frac{dU_{as}(x^+)}{dx^+} = H_{as}(x^+)U_{as}(x^+)
\end{equation}
can then be used to generate an asymptotic space
\begin{equation}
{\cal H}_{as} = exp[-\Omega^A(x^+)]{\cal H}_F
\end{equation}
from the usual Fock space ${\cal H}_F$, in the limit $x^+ \rightarrow -\infty$, where
$\Omega^A(x^+)$ is the asymptotic evolution operator defined by 
\begin{equation} 
U_{as}(x^+) = exp[-iH_0x^+] exp[\Omega^A(x^+)]
\end{equation}
The asymptotic evolution operator is then used to define the coherent states
\begin{equation}
\vert n : coh \rangle = exp[-\Omega^A] \vert n \rangle
\end{equation} 
The method of asymptotic dynamics, proposed originally by Kulish and Faddeev~\cite{KUL70} in the context of equal time QED consists of 
identifying the terms in the interaction Hamiltonian which do not vanish at infinitely large times and then using them to construct the 
asymptotic M\"oller operator and hence the coherent states. 
In LFFT's, the true IR divergences corresponding to $k^+,~{\bf k}_\perp \rightarrow 0$ are not expected to appear when one uses this 
CSB to calculate the transition matrix elements. In the following sections, we will prove this statement to all orders for fermion mass renormalization in LFQED. 

Interaction Hamiltonian of LFQED in light front gauge is given by \cite{MUS91} 
\begin{displaymath}
H_I(x^+)=V_1(x^+)+V_2(x^+)+V_3(x^+)
\end{displaymath}
where
\begin{equation}\label{V_1}
V_1(x^+)= e\sum_{i=1}^4 \int d\nu_i^{(1)}[ e^{-i \nu_i^{(1)} x^+} {\tilde h}_i^{(1)}(\nu_i^{(1)})
+ e^{i \nu_i^{(1)} x^+}
{\tilde h}^{(1)\dagger}_i (\nu_i^{(1)})]
\end{equation}
${\tilde h}^{(1)}_i(\nu^{(1)}_i)$ and $\nu_i^{(1)}$ are three point QED interaction vertex and the light front energy transferred at the vertex ${\tilde h}^{(1)}$ respectively. 
$V_2$ and $V_3$ are the non-local 4--point instantaneous vertices.
Here, we will focus on construction of asymptotic Hamiltonian using 3--point vertex. The same procedure can be used to construct the asymptotic Hamiltonian 
terms corresponding to the 4--point interaction vertices also. The details of the calculation of the four point asymptotic interaction can be found in Ref.~\cite{JAI12}. 
One can notice, from the time dependence of $V_1(x^+)$ that it does not become zero at large (light-cone) times whenever 
$\nu_i^{(1)}= 0$. For example, the three point interaction Hamiltonian has a term with light-cone time dependence of the form $exp[-i(p^- -k^- - (p-k)^-)x^+]$ and therefore, if  
$(p^-- k^- - (p-k)^-)\rightarrow 0$ then this term does not vanish at infinite times. Thus, the asymptotic Hamiltonian has a contribution from this term which can be written as 
\begin{align}\label{V1as}
V_{1as}(x^+) = e \sum_{i=1,4} \int d\nu_i^{(1)} \Theta_\Delta(k)[ e^{-i \nu_i^{(1)} x^+} \tilde h_i^{(1)}(\nu_i^{(1)})
+ e^{i \nu_i^{(1)} x^+}\tilde h^ {\dagger}_i (\nu_i^{(1)})] \;
\end{align}
where  $\Theta_\Delta(k)$ is the region of momentum space in which the energy difference $(p^-- k^- - (p-k)^-)$ becomes vanishingly small.
It can be shown that this condition is satisfied in the region defined by 
\begin{displaymath}
{\bf k}_\perp ^2 < {{k^+ \Delta} \over{p^+}}, \quad k^+ < {{p^+ \Delta} \over {m^2}}\;.
\end{displaymath}
We call this region the asymptotic region.
$\Theta_\Delta(k)$ in Eq.(8) is given by 
$$ \Theta_\Delta(k)=\theta\bigg({{k^+\Delta} \over p^+} - {\bf k}_\perp^2\bigg)
\theta\bigg({{p^+\Delta} \over m^2} - k^+\bigg)$$

Substituting $k^+\rightarrow 0$, ${\bf k_\perp}  \rightarrow 0 $ in all slowly varying functions of k and performing the 
$x^+$ integration, one obtains the asymptotic M\"oller operator which gives the asymptotic states as
\begin{align}
\Omega_{\pm}^A \vert n\colon p_i& \rangle=exp\biggl[-e\int{dp^+d^2{\bf p}_\perp}\sum_{\lambda=1,2}[d^3k][f(k,\lambda:p) 
a^\dagger(k,\lambda)- f^*(k,\lambda:p)a(k,\lambda)]\nonumber\\
&+e^2\int{dp^+d^2{\bf p}_\perp} \sum_{\lambda_1, \lambda_2=1,2}[d^3k_1][d^3k_2][g_1(k_1,k_2,\lambda_1,\lambda_2 \colon p) 
a^\dagger(k_2,\lambda_2)a(k_1,\lambda_1)-\nonumber\\
&g_2(k_1,k_2,\lambda_1,\lambda_2 \colon p)a(k_2,\lambda_2)a^\dagger(k_1,\lambda_1)]
\rho(p)\biggr]\vert n \colon p_i \rangle
\end{align}
where 
\begin{align}\label{theta2}
f(k,\lambda \colon p) = &{{p_\mu\epsilon_\lambda^\mu(k)} \over {p\cdot k}}\theta\bigg(\frac{k^+\Delta}{p^+}-{\bf k}_\perp^2\bigg) \theta\bigg(\frac{p^+\Delta}{m^2}-k^+\bigg) \;,\nonumber\\
f(k,\lambda \colon p) = &f^*(k, \lambda \colon p) \;,
\\
g_1(k_1,k_2,\lambda_1,\lambda_2 \colon p)=&-\frac{4p^+}{p \cdot k_1-p \cdot k_2+k_1 \cdot k_2} \delta^{3}(k_1-k_2)\nonumber\\
g_2(k_1,k_2,\lambda_1,\lambda_2 \colon p)=&\frac{4p^+}{p \cdot k_1-p \cdot k_2-k_1 \cdot k_2}\delta^{3}(k_1-k_2)
\end{align}
The second term here arises from the 4--point instantaneous interaction, which we have not included in Eq.~\ref{V1as}. Following the same procedure as in Ref.~\cite{ANU94}, 
we have used these asymptotic states to calculate the transition matrix elements and to demonstrate 
the absence of IR divergences in fermion self energy correction up to two loop level \cite{JAI12, JAI13}. In this work, we present an all 
order proof of cancellation of IR divergences in fermion self energy correction in CSB using the method of induction. 

\section{Graphical Representation of IR Finite Diagrams}\label{blob}

In  light-front time ordered perturbation theory, the transition matrix is given by the perturbative expansion
\begin{equation}\label{T}
T= V + V {1 \over {p^--H_0}}V + \cdots
\end{equation}
The electron mass shift is obtained by calculating $T_{pp}$ which is the matrix element of the above series 
between the initial and the final electron states $\vert p,s \rangle$  and it is 
given by,   
\begin{align}\label{deltam2}
\delta m^2= p^+ \sum_{s} T_{pp}
\end{align} 
\vskip -0.5cm
where 
\begin{align}
T_{pp}=\langle p,s \; \vert \; T \;\vert\; p,s \rangle =T^{(1)}+T^{(2)}+\cdots
\end{align}
In general, $T^{(n)}$ gives the $O(e^{2n})$ contribution to lepton self energy correction. 
The strategy used to develop a proof of cancellation of IR divergences to all orders in LFQED is based on the 
method of induction. To begin with, we have shown the cancellation of IR divergences up to $O(e^4)$ using CSB in LF gauge \cite{JAI12}. 
Now, we assume that the IR divergences cancel up to $O(e^{2n})$ and we  represent the 
$O(e^{2n})$ IR finite amplitude by a blob. A blob represents the sum of the Fock and coherent state contributions to the  self energy 
correction which give IR finite amplitude. The blob is of $O(e^{2n})$ and contains a maximum of n photon lines or    
2n 3--point interaction vertices. In case of diagrams containing 4-point instantaneous interaction vertices or having soft photon insertions, these 
numbers will be less than n and 2n respectively. Then the final task would be to express the $O(e^{2(n+1)})$ contributions in terms of this blob and show the cancellation of IR divergences in 
$O(e^{2(n+1)})$ in CSB. The general expression for transition matrix element in $O(e^{2n})$ is a sum of terms of the form:
\begin{align}\label{Tn}
T^{(n)}_j&=-\frac{e^{2n}}{2p^+ (2\pi)^{3n}}\int \prod_i\frac{d^3k_i}{2k_i^+ 2p_{2i-1}^+}\nonumber\\
 &\times\frac{\overline{u}(p,s)\epsilon\llap/_1(p\llap/_1+m)\epsilon\llap/_2(\slashed p_2+m)\cdots \, \cdots \cdots(p\llap/_\ell+m)\epsilon\llap/_\ell u(p, s)}
 {\displaystyle \prod_r (p^--p_r^--\sum_i k_i)}
\end{align}
where $r$, $i$ and $\ell$ depend on the kind of diagram $T^{(n)}_j$ represents. This enables us to express the $O(e^{2n})$  contribution as 
\begin{align}\label{T}
 T^{(n)}=\sum_j T^{(n)}_j=\sum_j \frac{\overline{u}(p,s)\mathcal{M}_n^{(j)} u(p,s)}{\mathcal{D}^{(j)}}
\end{align}
where $j$ is summed over all possible diagram in $O(e^{2n})$. $T^{(n)}$ will be assumed to be IR finite. Here
\begin{equation}
\mathcal{D}^{(j)}=\displaystyle \prod_r \mathcal{D}_r^{(j)} 
\end{equation}
and $\mathcal{D}^{(j)}$ corresponds to the product of  all the energy denominators, $\mathcal{D}_r^{(j)}$, corresponding to $r$ different intermediate states in $j^{th}$ diagram.
\begin{figure}
\centering
 \includegraphics[width=21cm,height=3cm,keepaspectratio]{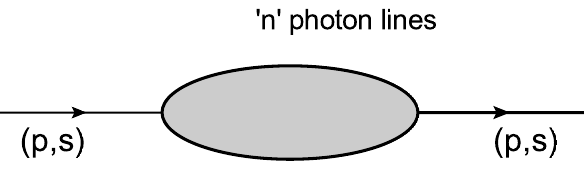}
 \caption{Basic diagram representing the sum of all diagrams of $O(e^{2n})$ in Fock state and CSB}
 \label{blob1}
\end{figure}
The graphical representation of the blob is shown in Fig.~\ref{blob1}. $ T^{(n)}$  will be assumed  to be IR divergence free. 
\section{An Example: Cancellation of Infrared divergences up to fourth order}
Before presenting the all order proof, we will first revisit the proof of cancellation of IR divergences in $\delta m^2$ up to $O(e^{4})$ to illustrate our strategy. 
In Ref.~\cite{JAI12, JAI13} we showed that the true infrared divergences  in $\delta m^2$  get cancelled up to $O(e^4)$ if one uses CSB instead of Fock basis to calculate the transition matrix elements \cite{JAI12}.
Now we sketch the proof for the same based on graphical  method which will be generalized to all orders in the next section. For the 
sake of simplicity, we will consider the contribution due to the 3$-$point 
interaction terms only.
\begin{figure}[h]
\centering
 \includegraphics[width=15cm,height=3cm,keepaspectratio]{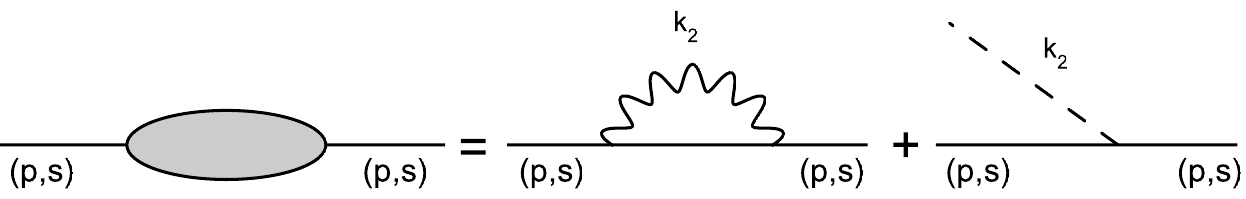}
 \caption{IR finite $O(e^2)$ blob which represents the sum of Fock state and coherent state contributions}
 \label{Oe2blob}
\end{figure}

To begin with, consider the  $O(e^{2})$ corrections which are represented by the two diagrams on the r.h.s. of Fig.~\ref{Oe2blob}. It has been shown in Ref.~\cite{JAI12, JAI13} that the sum of these two is  IR finite. 
In our notation, it is equal to 
\begin{equation} 
T^{(2)} = \sum_j \frac{\overline{u}(p,s)\mathcal{M}_2^{(j)} u(p,s)}{\mathcal{D}^{(j)}}
\end{equation} 
where the sum runs over 1 and 2 corresponding  to the two diagrams. The sum is represented by the IR finite blob on the l.h.s. in Fig.~\ref{Oe2blob}.

\begin{figure}[h]
\centering
 \includegraphics[width=36cm,height=3cm,keepaspectratio]{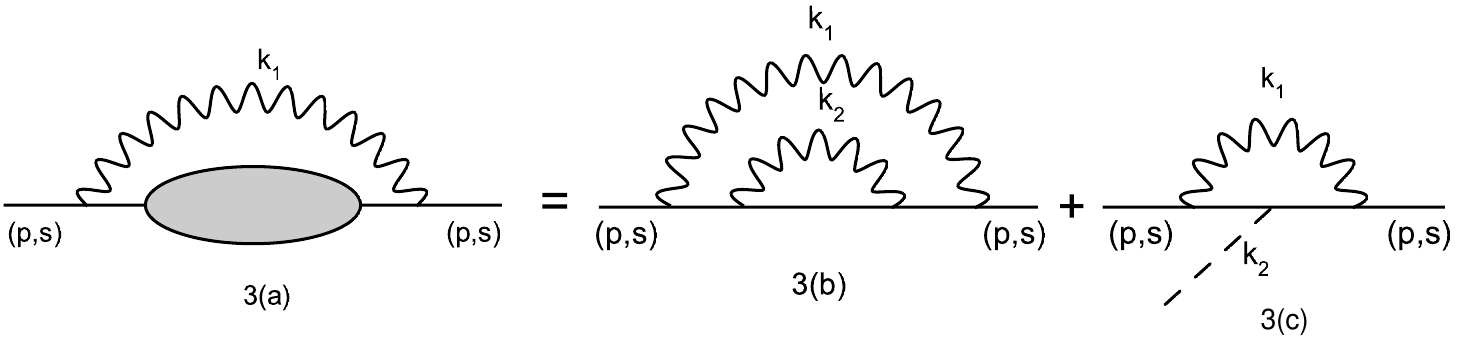}
 \caption{IR finite $O(e^2)$ blob with an external photon line results into $O(e^4)$ diagram in Fock basis}
 \label{Order4}
\end{figure}

Now consider the two diagrams on the r.h.s. of Fig.~\ref{Order4}, which are actually 
Fig. 3(a) and 8(b) in Ref.~\cite{JAI12} and have  been shown to be equal to 
\begin{align}\label{T_3a}
T_{3b}^{(2)}&={e^4\over{(2 \pi)^6}}\int{{d^2 {\bf k}_{1\perp}}{d^2{\bf k}_{2\perp}}\over 2p^+ } \int{{dk_1^+}{dk_2^+}\over{32 k_1^+k_2^+p_1^+p_2^+p_3^+}}\nonumber\\
&\times\frac{\overline u(p,s)[\slashed\epsilon^{\lambda_1}(k_1)(\slashed p_1+m)\slashed\epsilon^{\lambda_2}(k_2)(\slashed p_2+m)\slashed\epsilon^{\lambda_2}(k_2)(\slashed p_1+m)\slashed\epsilon^{\lambda_1}(k_1)]u(p,s)} {(p^--p_1^--k_1^-)(p^--p_2^--k_1^--k_2^-)(p^--p_1^--k_1^-)}
\end{align}
and
\begin{align}\label{T_3c}
T_{3c}^{(2)}&=\frac{e^4}{(2\pi)^6}\int\frac{d^2{\bf k}_{1\perp}d^2{\bf k}_{2\perp}}{2p^+}\int\frac{dk_1^+ dk_2^+}
{16 k_1^+k_2^+p_1^+p_2^+}\nonumber\\&
\times\frac{\overline u(p,s)[\slashed\epsilon^{\lambda_1}(k_1)(\slashed p_1+m) 
\slashed\epsilon^{\lambda_2}(k_2)(\slashed p_1+m)\slashed\epsilon^{\lambda_1}(k_1)]u(p,s)(p\cdot\epsilon^{\lambda_2}(k_2)) 
\Theta_\Delta (k_2)}{(p\cdot k_2)(p^--p_1^-- k_1^-)(p^--p_2^-- k_1^--k_2^-)}
\end{align}
respectively. Here, $p_1=p-k_1$ and $p_2=p-k_1-k_2$. The sum of Eqs.~(\ref{T_3a}) and (\ref{T_3c}) in our new notation is, 
\begin{align}\label{T3}
T_{3a}^{(2)}&=T_{3b}^{(2)}+T_{3c}^{(2)}\nonumber \\
	    &=\frac{e^2}{(2\pi)^3}\int \frac{d^3k_1}{2k_1^+}\frac{\overline{u}(p, s)\epsilon\llap/(k_1)(\slashed p_1+m)\mathcal{M}_2^{(j)} (\slashed p_1+m) \epsilon\llap/(k_1) u(p,s)}
	      {(p\cdot k_1)^2 \mathcal{D}^{(j)}}
 \end{align}
Note that the l.h.s. of Fig.~\ref{Oe2blob} minus the external lines is $\mathcal{M}_2^{(j)}$ for the jth diagram and hence the sum of the two $O(e^4)$ diagrams
on the r.h.s. of Fig.~\ref{Order4} is represented by Eq.~(\ref{T3}). 
\begin{figure}[h]
\centering
\includegraphics[width=30cm,height=3cm,keepaspectratio]{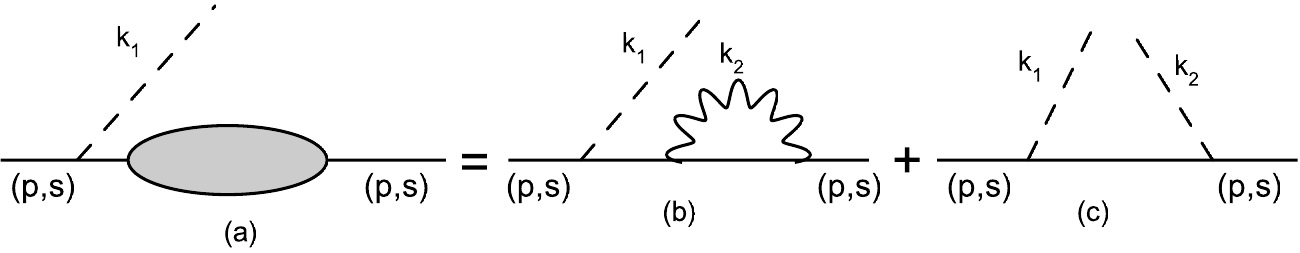}
\caption{IR finite $O(e^2)$ blob with an external photon line results into $O(e^4)$ diagram in coherent basis}
\label{fig4}
\end{figure}
Since the blob is IR finite, the IR divergences can appear ``only'' from the vanishing energy denominators of the kind 
$(p^--k_1^--(p-k_1)^-)$. All the other energy denominators inside the blob, which depend on $k_1$ and $k_2$ need not be taken into account.

\begin{figure}[h]
\centering
\includegraphics[width=30cm,height=5cm,keepaspectratio]{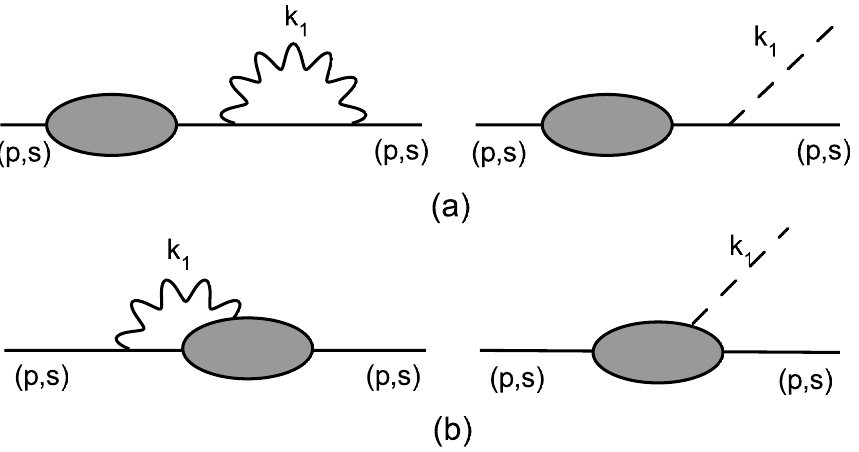}
\caption{Additional  diagrams corresponding to $O(e^4)$ contributions in Fock basis and CSB.}
\label{fig5}
\end{figure}

Similarly,  in CSB, the r.h.s of Fig.~\ref{fig4} is Fig.~8(a) and 9(d) of Ref.~\cite{JAI12} and which were shown to be equal to 
\begin{align}\label{4a}
T_{4b}^{(2)}&={e^4\over{(2 \pi)^6}}\int{{d^2 {\bf k}_{1\perp}}{d^2{\bf k}_{2\perp}}\over 2p^+ } \int{{dk_1^+}{dk_2^+}\over{32 k_1^+k_2^+p_1^+p_2^+p_1^+}}\nonumber\\
&\times\frac{\overline u(p,s)[\slashed\epsilon^{\lambda_2}(k_2)(\slashed p_2+m)\slashed\epsilon^{\lambda_2}(k_2)(\slashed p_1+m)\slashed\epsilon^{\lambda_1}(k_1)]u(p,s)(p\cdot \epsilon(k_1))\Theta_\Delta(k_1)}
{(p^--p_1^--k_1^-)(p^--p_3^--k_2^-)(p^--p_1^--k_1^-)}
\end{align}
and
\begin{align}\label{4c}
T_{4c}^{(2)}&=-{e^4\over{(2 \pi)^6}}\int{{d^2 {\bf k}_{1\perp}}{d^2{\bf k}_{2\perp}}\over 2p^+ } \int{{dk_1^+}{dk_2^+}\over{32 k_1^+k_2^+p_1^+p_2^+p_1^+}}\nonumber\\
&\times\frac{\overline u(p,s)[\slashed\epsilon^{\lambda_2}(k_2)(\slashed p_1+m)\slashed\epsilon^{\lambda_1}(k_1)]u(p,s)(p\cdot \epsilon(k_1))(p\cdot \epsilon(k_2))\Theta_\Delta(k_1)\Theta_\Delta(k_2)} 
{(p^--p_1^--k_1^-)(p^--p_3^--k_1^-)(p^--p_3^--k_2^-)}
\end{align}
respectively. Here, $p_3=p-k_2$. 
An additional contribution in CSB is obtained by adding  Eqs.~(\ref{4a}) and (\ref{4c}) and will be represented in our new notation as 
\begin{align}\label{T4}
T_{4a}^{(2)}=-\frac{e^2}{(2\pi)^3}\int \frac{d^3k_1}{2k_1^+}\frac{\overline{u}(p, s)\mathcal{M}_2^{(j)}(\slashed p_1+m)\epsilon\llap/(k_1) u(p,s)(p\cdot k_1)}
 {(p\cdot k_1)^2 \mathcal{D}^{(j)}}
 \end{align}
 and is represented graphically by the l.h.s. of Fig.~\ref{fig4}, while the r.h.s. is the sum of Eqs.~(\ref{4a}) and (\ref{4c}). 
In the limit, $k_1^+ \rightarrow 0, {\bf k}_{1\perp} \rightarrow 0$, the numerator of Eqs.~(\ref{T3}) and (\ref{T4}) become equal and the sum of their contributions is IR finite. 
At $O(e^4)$, there are two other diagrams involving only the three point vertices which contribute to self energy correction.  A similar argument can be 
constructed for the remaining two diagrams as well. The cancellation of IR divergences of these remaining diagrams is illustrated graphically  in Fig.~\ref{fig5}. 
\section{Cancellation of Infrared divergences to all orders}
Now, we consider an $O(e^{2n})$  blob which we will assume to be free of IR divergences. 
\begin{figure}[h]
\centering
\includegraphics[width=30cm,height=7cm,keepaspectratio]{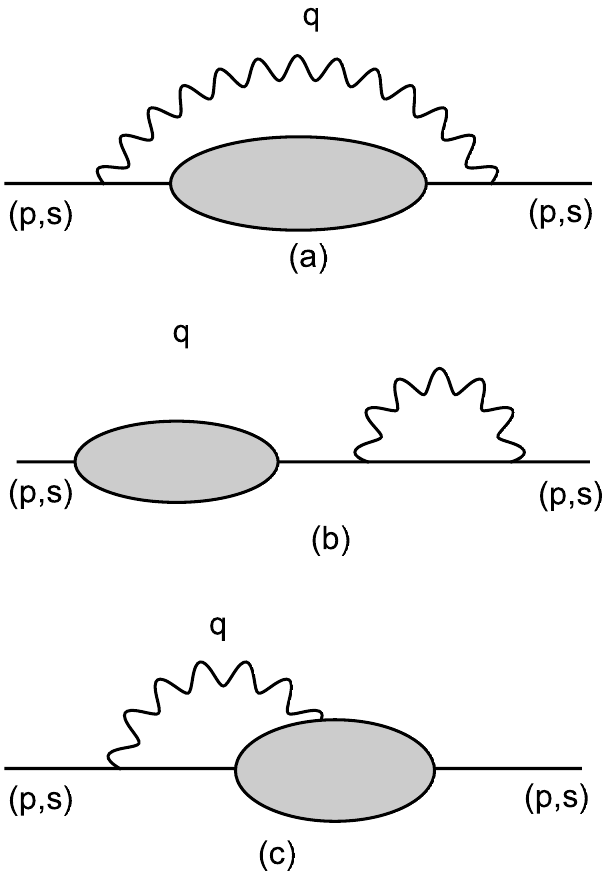}
 \caption{Addition of a photon line to $O(e^{2n})$ blob in Fock basis}
\label{blobn2}
\end{figure}
We shall show that the cancellation of IR divergences in $O(e^{2(n+1)})$ contribution to fermion mass renormalization in LFQED follows from this assumption. 
To construct an $O(e^{2(n+1)})$ diagram in Fock basis, we can add a photon to $n^{th}$ order blob in three different ways as shown in Fig.~\ref{blobn2}.
The contributions coming from the diagram in Figs.~\ref{blobn2}(a) and (b), in the limit $q^+ \rightarrow 0, {\bf q_\perp} \rightarrow 0$   are given by 
\begin{align}\label{T6a}
 T^{(n+1)}_{6a}=&\frac{e^2}{(2\pi)^3}\int \frac{d^3q}{2q^+}\frac{\overline{u}(p,s)\epsilon\llap/(q)(\slashed P+m)\mathcal{M}_n^{(j)} (\slashed P+m) \epsilon\llap/(q) u(p,s)}
 {(p\cdot q)^2 \mathcal{D}^{(j)}}\\ \label{T6b}
 T^{(n+1)}_{6b}=&-\frac{e^2}{(2\pi)^3}\int \frac{d^3q}{2q^+}\frac{\overline{u}(p,s)\epsilon\llap/(q)(\slashed P+m)\epsilon\llap/(q) (\slashed p^\prime+m) \mathcal{M}_n^{(j)} u(p,s)}
 {(p\cdot q)(p^--p^{\prime-})\mathcal{D}^{(j)}}  
%  T^{(n+1)}_{6c}=&\frac{e^2}{(2\pi)^3}\int \frac{d^3q}{2q^+}\frac{\overline{u}(p,s)\mathcal{M}_n^{(j)} (\slashed P+m)\epsilon\llap/(q) u(p,s)}
%  {(p\cdot q)\mathcal{D}^{(j)}}
\end{align}
where $P=p-q$ and  $p^\prime = p$.
\begin{figure}[h]
\centering
\includegraphics[width=30cm,height=7cm,keepaspectratio]{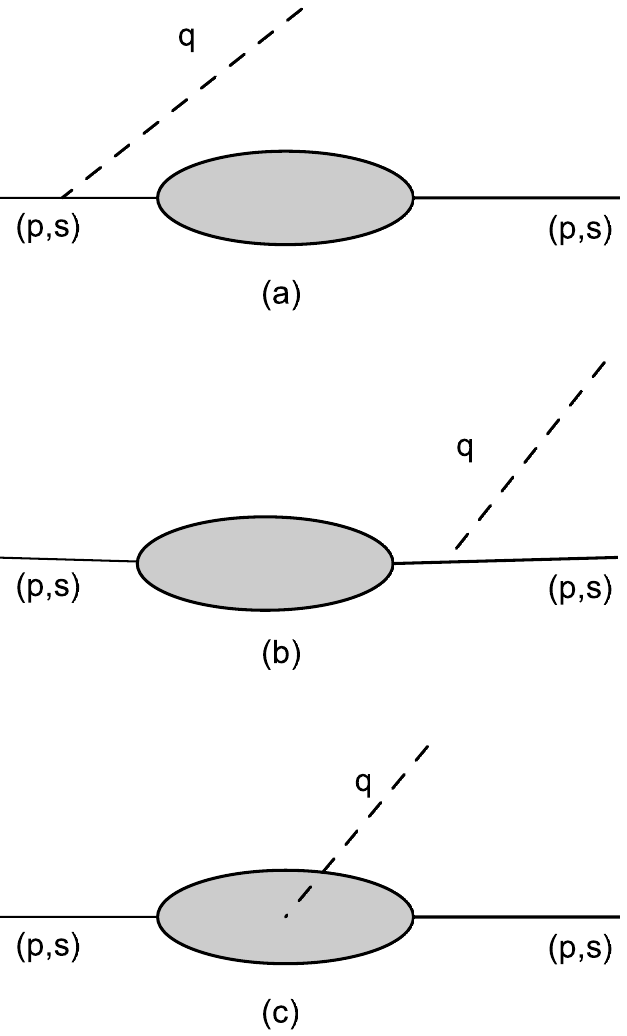}
\caption{Addition of a photon line to $O(e^{2n})$ blob in CSB}
\label{blob3}
\end{figure}
The additional contributions in $(n+1)^{th}$ order, when we use the CSB, are  shown  in Fig.~\ref{blob3} and are  given by
\begin{align}\label{T7a}
 T^{\prime(n+1)}_{7a}=&-\frac{e^2}{(2\pi)^3}\int \frac{d^3q}{2q^+}\frac{\overline{u}(p,s)\mathcal{M}_n^{(j)}(\slashed P+m)\epsilon\llap/(q) u(p,s)(p \cdot \epsilon(q))\,\Theta_\Delta(q)}
 {(p\cdot q)^2\mathcal{D}^{(j)}}\\ \label{T7b}
 T^{\prime(n+1)}_{7b}=&\frac{e^2}{(2\pi)^3}\int \frac{d^3q}{2q^+}\frac{\overline{u}(p,s)\epsilon\llap/(q) (\slashed p^\prime+m) \mathcal{M}_n^{(j)} u(p,s)(p\cdot \epsilon(q))\,\Theta_\Delta(q)}
 {(p\cdot q)(p^--p^{\prime-})\mathcal{D}^{(j)}}
%  T^{\prime(n+1)}_{7c}=&-\frac{e^2}{(2\pi)^3}\int \frac{d^3q}{2q^+}\frac{\overline{u}(p,s)\mathcal{M}_n^{(j)} u(p,s)(p \cdot \epsilon(q))\,\Theta_\Delta(q)}{(p\cdot q)\mathcal{D}^{(j)}}
\end{align}

In the limit, $q^+\rightarrow 0$, ${\bf q}_\perp \rightarrow 0$, 
$(\slashed P+m)\epsilon\llap/(q) u(p,s) \rightarrow (p \cdot \epsilon(q)) u(p,s)$. 
As a result, the sum of Eqs.~(\ref{T6a}) and (\ref{T6b}) exactly cancels the sum of Eqs.~(\ref{T7a}) and (\ref{T7b}). 
i.e. the IR divergences in Figs.~\ref{blobn2}(a) and \ref{blobn2}(b) are exactly cancelled by those in Figs.~\ref{blob3}(a) and \ref{blob3}(b).
\section{Overlapping diagrams}
The overlapping diagrams as in Figs.~\ref{blobn2}(c) and \ref{blob3}(c) need a special treatment as in this case attaching a photon line inside the blob changes the structure of the numerator and denominators and thus the blob presented for calculating Figs. \ref{blobn2}(a), \ref{blobn2}(b), \ref{blob3}(a) and \ref{blob3}(b) in the previous section  is different from the one shown in Figs.~\ref{blobn2}(c) and \ref{blob3}(c). 
It can be shown that when a photon is absorbed by an IR finite blob, we need to consider only the IR divergences in the limit $q^+\rightarrow 0$, ${\bf q}_\perp \rightarrow 0$. The general structure of an overlapping diagram is of the form 
\begin{align}\label{T6c}
 T^{(n+1)}_{6c}=&\frac{e^2}{(2\pi)^3}\int \frac{d^3q}{2q^+}\frac{\overline{u}(p,s)\mathcal{M}_n^{\ell(j)} (\slashed P+m)\epsilon\llap/(q) u(p,s)}
 {(p\cdot q)\mathcal{D}^{(j)}}
\end{align}
where $\mathcal{M}_n^{\ell(j)}$ is $\mathcal{M}_n^{(j)}$ with  an extra $q$ attached at the $\ell^{th}$ vertex inside the blob as shown in Fig.~\ref{blobn2}(c). $\mathcal{M}_n^{\ell(j)}$ can be written explicitly as;
\begin{align}
 \mathcal{M}_n^{\ell(j)}= \epsilon\llap/(k_1)(\slashed P_1+m)\epsilon\llap/(k_2)(\slashed P_2+m)\cdots \cdots \epsilon\llap/(k_\ell)(P_\ell+m)\epsilon\llap/(q) \epsilon\llap/(p_{\ell+1}+m)\cdots \cdots \cdots
\end{align}
The energy denominators corresponding to the intermediate states will be
\begin{align}
 \mathcal{D}^{(j)}= (p^--p_1^-- k_1^--q^-)(p^--p_2^-- k_1^--k_2^--q^-)\cdots \cdots (p^--p_{\ell}^-- \sum_i k_i^--q^-) \cdots \cdots\cdots 
\end{align}
The coherent state diagram corresponding to the case in which $q$ is emitted  at $\ell^{th}$ vertex in the blob  is given by
\begin{align}
 \label{T7c}
 T^{\prime(n+1)}_{7c}=&-\frac{e^2}{(2\pi)^3}\int \frac{d^3q}{2q^+}\frac{\overline{u}(p,s)\mathcal{M}_n^{(j)} u(p,s)(p \cdot \epsilon(q))\,\Theta_\Delta(q)}{(p\cdot q)\mathcal{D}^{(j)}}
\end{align}
$T^{\prime(n+1)}_{7c}$ gives equal and opposite contribution to Eq.~(\ref{T6c}). There can be additional $(n-1)$ such overlapping diagrams both in fock and corresponding coherent state bases. 
Each of the fock state contribution cancels the corresponding coherent state basis contribution at $O(e^{2(n+1)})$ i.e.
the sum of Eq.~(\ref{T6c}) and (\ref{T7c}) is infrared finite.

It is to be noted that even if the blob contains 4--point instantaneous vertices, Figs. 6 and 7 are the only diagrams in next order in graphical notation and therefore, the 
argument presented in this section still holds. 

\section{Conclusion}
We have demonstrated,  using the  coherent state formalism,   that the true IR divergences in self energy correction cancel to all order in 
LFQED. The proof presented here can be extended to a general n-point amplitude. 
We plan to address this problem in future.  

It is well known that in QCD the IR divergences are not cancelled. One can apply the same procedure as in QED to QCD as shown in \cite{ANU00, BUT78}  but then one must 
obtain the coherent states which incorporate large distance limit of QCD potential. As we know that, in QCD, the in and out states are nothing but the bound states of quarks and gluons, hence one needs to use the method of  asymptotic dynamics carefully. 
One must add to the free Hamiltonian not just the large distance limit of QCD potential but also the confining potential which arises due to the bound state of quarks and gluons in the hadrons. This is the reason why the coherent state formalism was  found to be insufficient for the cancellation of IR divergences in  higher orders in equal time QCD.
The non-cancellation of IR divergences in QCD turns out to be an interesting puzzle, which if solved will help to understand the color confinement properties in QCD. We hope that form of artificial confining potential is the key to the puzzle and will also help us to device an all order proof in QCD. 
We hope that the coherent state formalism may provide a hint towards constructing an artificial potential for QCD which  can be used to perform bound state calculations\cite{WILSON94}.

 This is due to the fact that the asymptotic states in strong interactions are 
bound states and not just the coherent states as in QED.  One way to obtain  the "artificial"  confining potential required for light front bound state calculations \cite{WILSON94} 
would be to impose the criteria of IR cancellation to find an appropriate potential \cite{ANU05}. The techniques developed in present work are expected to play an important role in this. 
\section{Acknowledgement}
AM would like to thank BRNS for financial support  under the Grant No. 2010/37P/47/BRNS and JM would like to thank Department of Physics, Mumbai University.


\begin{thebibliography}{abcdef99}
\bibitem{WILSON94} 
  K.~G.~Wilson, T.~S.~Walhout, A.~Harindranath, W.~-M.~Zhang, R.~J.~Perry and S.~D.~Glazek,
  %``Nonperturbative QCD: A Weak coupling treatment on the light front,''
  Phys.\ Rev.\ D {\bf 49}, 6720 (1994)
  [hep-th/9401153].
\bibitem{BRODSKY97} 
  S.~J.~Brodsky, H.~-C.~Pauli and S.~S.~Pinsky,
  %``Quantum chromodynamics and other field theories on the light cone,''
  Phys.\ Rept.\  {\bf 301}, 299 (1998)
  [hep-ph/9705477].
\bibitem{BLOCH37} F.~Bloch and A.~Nordsieck,
  %``Note on the Radiation Field of the electron,''
  Phys.\ Rev.\  {\bf 52}, 54 (1937).
\bibitem{YENNIE61} D.~R.~Yennie, S.~C.~Frautschi and H.~Suura,
  %``The infrared divergence phenomena and high-energy processes,''
  Annals Phys.\  {\bf 13}, 379 (1961). 
  \bibitem{CHU65} V.~Chung,
  %``Infrared Divergence in Quantum Electrodynamics,''
  Phys.\ Rev.\  {\bf 140}, B1110 (1965).
\bibitem{KUL70} P.~P.~Kulish and L.~D.~Faddeev,
  %``Asymptotic conditions and infrared divergences in quantum electrodynamics,''
  Theor.\ Math.\ Phys.\  {\bf 4}, 745 (1970).
%   [Teor.\ Mat.\ Fiz.\  {\bf 4}, 153 (1970)].% 
\bibitem{GRECO78}M.~Greco, F.~Palumbo, G.~Pancheri-Srivastava and Y.~Srivastava,
  %``Coherent State Approach to the Infrared Behavior of Nonabelian Gauge Theories,''
  Phys.\ Lett.\ B {\bf 77}, 282 (1978).
 \bibitem{Vary1988} A. Harindranath and J.P. Vary, Phys. Rev. D 37, 3010
(1988).
 \bibitem{Vary1999}L. Martinovic and J.P. Vary, Phys. Lett. B 459, 186
(1999).
\bibitem{ANU94}A.~Misra,
  %``Coherent states in null plane QED.,''
  Phys.\ Rev.\ D {\bf 50}, 4088 (1994)
  [hep-th/9311101].
\bibitem{ANU00}A.~Misra,
  %``Coherent states in light front QCD,''
  Phys.\ Rev.\ D {\bf 62}, 125017 (2000).
\bibitem{JAI12} J.~D.~More and A.~Misra,
  %``Infra-red Divergences in Light-Front QED and Coherent State Basis,''
  Phys.\ Rev.\ D {\bf 86}, 065037 (2012)
  [arXiv:1206.3097 [hep-th]] 
  %and references therein.
\bibitem{JAI13} J.~D.~More and A.~Misra,
  %``Fermion Self Energy Correction in Light-Front QED using Coherent State Basis,''
  Phys.\ Rev.\ D {\bf 87}, 085035 (2013)
  [arXiv:1302.3522 [hep-th]].
\bibitem{ANU96} A.~Misra,
  %``Light cone quantization and the coherent state basis,''
  Phys.\ Rev.\ D {\bf 53}, 5874 (1996).
  \bibitem{BUT78}D.~R.~Butler and C.~A.~Nelson,
  %``Nonabelian Structure of {Yang-Mills} Theory and Infrared Finite Asymptotic States,''
  Phys.\ Rev.\ D {\bf 18}, 1196 (1978); C.~A.~Nelson,
  %``Origin of Cancellation of Infrared Divergences in Coherent State Approach: Forward Process q q ---> q q Gluon,''
  Nucl.\ Phys.\ B {\bf 181}, 141 (1981); C.~A.~Nelson,
  %``Avoidance of Counter Example to Nonabelian Bloch-Nordsieck Conjecture by Using Coherent State Approach,''
  Nucl.\ Phys.\ B {\bf 186}, 187 (1981). 
\bibitem{MUS91}D.~Mustaki, S.~Pinsky, J.~Shigemitsu and K.~Wilson,
  %``Perturbative renormalization of null plane QED,''
  Phys.\ Rev.\ D {\bf 43}, 3411 (1991).
\bibitem{ANU05} A.~Misra,
  %``Method of asymptotic dynamics in light-front field theory,''
  Few Body Syst.\  {\bf 36}, 201 (2005).
\end{thebibliography}
\end{document}